# Elastic Spin-Hall Effect in Mechanical Graphene

Yao-Ting Wang and Shuang Zhang

*School of Physics and Astronomy, University of Birmingham, Birmingham B15 2TT, United Kingdom*

**We show that spin-orbit interaction and elastic spin-Hall effect can exist in a classical mechanical system consisting of a two-dimensional honeycomb lattice of masses and springs. The band structure shows the presence of splitting at K point induced by the difference of longitudinal and transverse elastic constant, and this splitting can be regarded as an effective Dresselhaus-type spin-orbit coupling. Interestingly, as an initial displacement away from the equilibrium is applied, the time evolution simulation shows that waves of different spin polarization propagates along different directions at the $\Gamma$ and K point, which is characteristic of spin-Hall effect. Several cases for spin-Hall effect are also discussed.**

## I. INTRODUCTION

Spin-orbit coupling (SOC) is an interaction between a particle's motion and spin, which accounts for the physical origin of the fine structure in early quantum mechanics development. In condensed matter physics, SOC plays very important role in understanding various interesting phenomena and generating diverse applications. One of the examples is that a strong spin-orbit interaction in ferromagnetic materials is able to generate a significant intrinsic magnetic field that gives rise to the anomalous Hall states. In the regime of semiconductors, SOC can be categorized into a generic symmetry-independent type, which exists in every material systems, and a symmetry-dependent type such as Rashba and Dresselhaus SOC. More recently, the latter type has drawn enormous attention as they lead to interesting phenomena in spintronics such as spin-Hall effect for weak interaction [1], anomalous Hall effect [2] and anomalous spin-Hall effect [3] for strongly coupled systems.

Inspired by the progresses of SOC related physics in electronic regime, effective SOC has been found in a variety of other physical systems including optics and acoustics. For example in photonics, by defining the left (right) circular polarization of light as spin up (down) state, the spin-Hall effect of light [4-6] has been proposed to describe a relative lateral shift between the two spin polarization states as light undergoes refraction or reflection at an interface between two different dielectric media. In addition, an effective intrinsic SOC realized by metamaterial structures has been put forward to mimic the Kane-Mele model in graphene [8] and to realize two-dimensional photonic topological insulators [7]. In phononics, similar spin-dependent splitting effect is also theoretically predicted for elastic waves are traveling through an interface [9]. Besides, the SOC concept has been theoretically predicted [10] and experimentally [11] verified for polaritons that exist in a quantum well embedded in a microcavity. Further, due to the fact that the photons with TE and TM polarizations are coupled inside microcavities [12], the dispersion diagram of a cavity-photonic crystal exhibits a Dresselhaus SOC at the K point[13].

The physics occurring in photonic systems can be intuitively linked to classical vibrating systems. Similar analysis of modes can be applied on the elastic waves or classical vibration since they naturally consist of longitudinal and transverse components. In this article, we introduce a mechanical graphene, namely, spring-mass model in honeycomb lattice, in which the longitudinal-transverse splitting gives rise to an effective intrinsic Dresselhaus type SOC. The proposed system opens up new perspectives towards manipulation of mechanical waves in artificial mechanical metamaterials.

One of the most intriguing phenomena induced by spin-orbit interaction is arguably the spin-Hall effect, which plays an important role in the research of spintronics. The spin-Hall effect is a transverse spin current driven by the electronic current, which could be caused by either intrinsic [14-15] or extrinsic SOC [16]. In this article, we introduce an analogy of spin-Hall effect based on an effective SOC of classical vibration. In what follows, we demonstrate a spin-wavevector correlated propagation for classical vibrations in a mechanical graphene, i.e. elastic spin-Hall effect (ESHE). Here we emphasize that this effect has a different physical mechanism in comparison to spin Hall effect of phonons [9], which describes the lateral displacements as a wave packet passes through an interface between two different elastic materials. In addition, we propose an equivalent system for elastic waves, which is made up of two isotropic elastic solids. The observed phonon dispersion and spin texture verify that it has the same signature of SOC.

## II.    MODEL

In order to introduce the ESHE, we first define "spin" as an in-plane rotation for the centre of mass of a rigid body, i.e. spin up/down for $x \pm iy$, which is also referred to "chiral phonons" in the recent literature [17]. Then we begin by discussing the spin-orbit interaction in a mechanical graphene. The system consists of a series of honeycomb-arranged rigid body spheres with mass $M$, and massless springs with longitudinal elastic constant $C_L$ and transverse one $C_T$, as shown in Fig. 1(a). Throughout the paper we assume that all the springs have good linearity and restrict ourselves to taking only in-plane vibrations into consideration. The distance between two nearest neighbour (NN) mass points (A and B) is $a$ and the lattice translation vectors are expressed as $\mathbf{r}_{mn} = m\mathbf{a}_1 + n\mathbf{a}_2$, where $\mathbf{a}_1 = \sqrt{3}a\hat{\mathbf{x}}$, $\mathbf{a}_2 = (\sqrt{3}\hat{\mathbf{x}} + \hat{\mathbf{y}})a/2$. The transverse elasticity arises from stretching of the springs, that is to say, the NN distance $a$ at equilibrium is longer than the natural length of the spring [18]. The three longitudinal and transverse unit vectors that connect between the nearest neighbours are

$$\hat{\mathbf{L}}_1 = (\sqrt{3}\hat{\mathbf{x}} + \hat{\mathbf{y}})/2, \; \hat{\mathbf{L}}_2 = (-\sqrt{3}\hat{\mathbf{x}} + \hat{\mathbf{y}})/2, \; \hat{\mathbf{L}}_3 = -\hat{\mathbf{y}},$$

$$\hat{\mathbf{T}}_1 = (-\hat{\mathbf{x}} + \sqrt{3}\hat{\mathbf{y}})/2, \; \hat{\mathbf{T}}_2 = -(\hat{\mathbf{x}} + \sqrt{3}\hat{\mathbf{y}})/2, \; \hat{\mathbf{T}}_3 = \hat{\mathbf{x}}.$$

By setting the displacements of the A and B as $\xi$ and $\eta$, respectively, the variation in length parallel and perpendicular to the spring orientation can be denoted as $\hat{\mathbf{L}}_i \cdot (\boldsymbol{\xi}_{mn} - \boldsymbol{\eta}_{mn})$ and $\hat{\mathbf{T}}_i \cdot (\boldsymbol{\xi}_{mn} - \boldsymbol{\eta}_{mn})$. Fig. 1(b) shows that vibrating mode transfer between the nearest neighbour only occurs between modes of the same vibration direction, whereas that between perpendicular vibration directions is prohibited. Following the preceding description, the

equation of motions for mechanical graphene is governed by

$$M\ddot{\xi}_{mn} = -C_L\left[\hat{L}_1\hat{L}_1\cdot(\xi_{mn}-\eta_{mn})+\hat{L}_2\hat{L}_2\cdot(\xi_{mn}-\eta_{m-1,n})+\hat{L}_3\hat{L}_3\cdot(\xi_{mn}-\eta_{m,n-1})\right]$$
$$-C_T\left[\hat{T}_1\hat{T}_1\cdot(\xi_{mn}-\eta_{mn})+\hat{T}_2\hat{T}_2\cdot(\xi_{mn}-\eta_{m-1,n})+\hat{T}_3\hat{T}_3\cdot(\xi_{mn}-\eta_{m,n-1})\right]$$ (1-a)

$$M\ddot{\eta}_{mn} = -C_L\left[\hat{L}_1\hat{L}_1\cdot(\eta_{mn}-\xi_{mn})+\hat{L}_2\hat{L}_2\cdot(\eta_{mn}-\xi_{m+1,n})+\hat{L}_3\hat{L}_3\cdot(\eta_{mn}-\xi_{m,n+1})\right]$$
$$-C_T\left[\hat{T}_1\hat{T}_1\cdot(\eta_{mn}-\xi_{mn})+\hat{T}_2\hat{T}_2\cdot(\eta_{mn}-\xi_{m+1,n})+\hat{T}_3\hat{T}_3\cdot(\eta_{mn}-\xi_{m,n+1})\right]$$ (1-b)

As studied previously [18], there exist four bands in the case $C_L \gg C_T$; two nearly flat bands of transverse modes at zero frequency and $\sqrt{3}\omega_0$, and two bands for longitudinal modes. On the other hand, for a system with $C_L$ slightly different from $C_T$, the energy of longitudinal (L) and transverse (T) are only slightly different, which implies an energy splitting owing to the LT discrepancy. To prove this argument, by setting $C_L = 4N/m$, and M =10g, we plot the frequency dispersions around the irreducible Brillouin zone in Fig. 2(a)-(c) for $C_T = C_L/8$, $C_L/3$ and $7C_L/8$. Note that even though the magnitude $7C_L/8$ might be unrealistically large, it helps on a clear explanation of SOC. The band structure for $C_T = C_L/8$ gives a Dirac degeneracy at K point with non-flat first and fourth bands, which is consistent with the result shown in ref. [19]. The gap at M between the 2[nd] and 3[rd] band closes at $C_T = C_L/3$ and it is observed that the dispersion is linear along direction Γ to M but parabolic along M to K. This type of dispersion can give rise to the trigonal warp effect [13,18,20], i.e. there exist three extra Dirac cones in the vicinity of K and K' points. These additional Dirac points are located at the high symmetry line Γ-M because of the three-fold symmetry. At $C_T = 7C_L/8$, band diagram illustrates a very similar splitting pattern to the bilayer graphene [21] or the graphene sheet including Rashba spin-orbit interaction [22].

In order to identify the nature of this SOC, by altering the original basis to circular bi-polarization basis $[\psi_A^\uparrow, \psi_A^\downarrow, \psi_B^\uparrow, \psi_B^\downarrow]^T$ with $\psi_{A(B)}^{\uparrow\downarrow}$ representing spinors $[\xi_x(\eta_x) \pm i\xi_y(\eta_y)]/\sqrt{2}$ at A and B lattice, equation (1) are rewritten to the Bloch Hamiltonian form as

$$\mathbf{H} = \begin{bmatrix} \frac{3C}{M} & 0 & \frac{C}{M}(1+s+p) & \frac{\Delta C}{M}\left[1 - \frac{s+p}{2} - \frac{i\sqrt{3}}{2}(s-p)\right] \\ 0 & \frac{3C}{M} & \frac{\Delta C}{M}\left[1 - \frac{s+p}{2} + \frac{i\sqrt{3}}{2}(s-p)\right] & \frac{C}{M}(1+s+p) \\ h.c. & h.c. & \frac{3C}{M} & 0 \\ h.c. & h.c. & 0 & \frac{3C}{M} \end{bmatrix}, \quad (2)$$

where $p = e^{i\mathbf{k}\cdot\mathbf{a}_1}$, $s = e^{i\mathbf{k}\cdot\mathbf{a}_2}$, $C = (C_L + C_T)/2$, and $\Delta C = (C_L - C_T)/2$. In equation (2), if we turn off $\Delta C$, which implies no LT coupling, the Hamiltonian expresses the feature as exactly the same as the graphene with a non-zero bias energy. As $\Delta C$ is not equal to zero, the splitting of phonon dispersion emerges as shown in Fig. 1. To confirm this splitting is caused from an effective SOC, we further plot the spin texture of the second band in the vicinity of K point Fig. 2(d) to (f). It is shown that, regardless of the magnitude of $C_T$, all the spin textures show similar characteristics to that of Dresselhaus SOC, which originally results from the inversion symmetry breaking in semiconductors. In the vicinity of K point $(\delta k_{x,y} a \ll 1)$, for a large transverse elastic constant one can derive a low-energy approximation of spin-orbit term in equation (2) that matches Dresselhaus SOC near K point as

$$H_D = \frac{3\Delta C}{4M}\left(\tau_z \sigma_x s_x - \sigma_y s_y\right) \quad (3)$$

where $\tau_z$ is the valley degree of freedom and $s_{x,y}$ represent the Pauli matrix for spins. Note that in equation (3) we have omitted $\delta k_{x,y} a$ and the higher terms, which makes it different from the result given by [13]. As the rotation motion of the mass represents spin polarization in mechanical graphene, there exists an effective Dresselhaus field in the system. In this analogue system, the length of arrows can be regarded as the strength of in-plane effective magnetic fields, which is a similar way as the real spins being aligned by real magnetic fields. The insets in Fig. 2(a) to (c) indicate the corresponding spin textures at Γ point. All three spin textures demonstrate nearly identical patterns indicating the robustness of the effective magnetic field against the change of $C_T$ at Γ point. In contrast, in the vicinity of K(K') points, the effective magnetic fields become weaker as $C_T$ decreases in Fig 2(d) to (f). This feature significantly affects the propagation of spin waves at the K(K') points, which will be discussed in more details in the next section.

## III.  ELASTIC SPIN HALL EFFECT

To figure out how the "spin current" evolves in a mechanical graphene, we apply a Gaussian pulse source as illustrated in Fig. 3. With time step Δt = 10 ms, Fig. 4(a)-(c) illustrate the evolution of spin fields for several high symmetry points (Γ, K, and K') after 100 time-step iterations. At Γ point, the corresponding Stokes' parameter $\rho_c = I_+ - I_-$, where $I$ is the intensity for two spins, is shown in figure. It is evident to see that the splitting for the four spin polarizations (spin-up in red and spin-down in blue colours) propagates radially in different directions because of the conservation of spin angular momentum. The field distribution shows a four-domain pattern as that of optical spin-Hall effect [10]. Fig. 4(b) and 4(c) show the ESHE at K and K' point, respectively. There are two spin polarizations in real space propagating in opposite direction, and it gives rise to a reverse splitting of circular polarization due to the field inversion at K and K' point.

As aforementioned, the transverse mode could be supported as long as the springs are stretched. Based on the result in ref. [18], one can define a factor $\alpha$ as the ratio between the spring's natural length and NN distance. The transverse elastic constant can be expressed as

$$C_T = C_L(1-\alpha) \qquad (4).$$

Despite the minimum value of $\alpha$ can reach to zero in principle, its magnitude is roughly limited within the region from 0.2 to 1 so as to keep the linearity of a spring. As such, the large transverse constant in the preceding discussion, $C_T = 7C_L/8$, requires an impractical system with the NN distance almost ten times to the original length of the spring. For a more practical system with a reasonable ratio α = 1/3, the corresponding $C_T$ is given by $2C_L/3$. Fig. 4(d)-(f) demonstrate the ESHE field pattern for α = 1/3 at Γ, K, and K' point after 100 time-step iteration. Fig. 4(d) exhibits similar spin propagation as that in the Fig. 4(a) for $C_T = 7C_L/8$. In fact, we discover that the ESHE at Γ point is not sensitive to the variation of α parameter. It is consistent with the preceding argument for the robustness against the alteration of $C_T$ at Γ point. Even for α further increased to 1/2, our simulation shows that the ESHE remains its four-domain spin-wave-splitting property and radial propagation. However, Fig. 4(e) and (f) indicate that the ESHE at K and K' point are distinct from Fig. 4(b) and (c), but they still fulfil the feature of field inversion. Yet, as shown in Fig. 2(d)-(e), the effective fields are weaker for small transverse elastic constant. For such weak field strength, it cannot give rise to an obvious splitting for two spin waves, such that the spin waves are mixed in Fig. 4(e) and (f).

## IV. CAVITY-PHONONIC CRYSTALS AS AN EQUIVALENT SYSTEM

Besides spring-mass systems, here we propose another equivalent elastic crystal system for realizing ESHE. The crystal is made by two isotropic elastic solids. In order to establish the equivalence between the mass-spring system and the elastic crystal, we first deduce the classical Hamiltonian for equation (1) in terms of the wave vector **k** as (See supplementary information for detailed derivation)

$$H = \frac{\left(\tilde{\Pi}_A^L\right)^2 + \left(\tilde{\Pi}_B^L\right)^2}{2M} + \frac{3C_L}{2}\left[\left(\tilde{L}_A\right)^2 + \left(\tilde{L}_B\right)^2\right] + \frac{\left(\tilde{\Pi}_A^T\right)^2 + \left(\tilde{\Pi}_B^T\right)^2}{2M} + \frac{3C_T}{2}\left[\left(\tilde{T}_A\right)^2 + \left(\tilde{T}_B\right)^2\right]$$
$$- \frac{C_L}{2}\left[\left(1 + p^* + s^*\right)\left(\tilde{L}_A\tilde{L}_B + \tilde{L}_B\tilde{L}_A\right)\right] - \frac{C_T}{2}\left[\left(1 + p^* + s^*\right)\left(\tilde{T}_A\tilde{T}_B + \tilde{T}_B\tilde{T}_A\right)\right], \quad (5)$$

where $\tilde{\Pi}_{A(B)}^{L(T)}$ and $\tilde{L}(T)_{A(B)}$ are the canonical momentum and displacement for L and T at A or B lattice, respectively. Eq. 4 shows that the local vibration for each mass point can be regarded as a simple harmonic oscillator in the presence of an elastic potential $\frac{3}{2}Cu^2$. Besides, the negative coupling term represents the energy transfer from one oscillator to another. The coupling coefficient plays the role of the hopping parameter of nearest neighbours in tight-binding description. For the sake of simplicity, one can neglect the energy of local simple harmonic oscillators since it only provides a bias and does not affect the main feature of phonon dispersion. In the basis of creation and annihilation operators, the Hamiltonian of a mechanical graphene reads

$$H = -\hbar t_L \left[ a^\dagger_{L,mn} b_{L,mn} + a^\dagger_{L,mn} b_{L,m-1,n} + a^\dagger_{L,mn} b_{L,m,n-1} + \text{H.c.} \right] \\ - \hbar t_T \left[ a^\dagger_{T,mn} b_{T,mn} + a^\dagger_{T,mn} b_{T,m-1,n} + a^\dagger_{T,mn} b_{T,m,n-1} + \text{H.c.} \right], \quad (6)$$

where the hopping parameter $t_{L(T)} = \omega_{L(T)}/3$ for L and T was introduced. Based on equation (6), we can extend the ESHE to elastic waves propagating inside a phononic crystal made by embedding tungsten carbide ( $\rho$ = 13800 kg/m³, $c_l$ = 6655 m/s, $c_t$ = 3980 m/s ) rods in an aluminium ( $\rho$ = 2690 kg/m³, $c_l$ = 6420 m/s, $c_t$ = 3040 m/s ) background. The dispersive relations in Fig. 5 are numerically calculated by COMSOL Multiphysics 5.1, a commercial package using finite-element method. In Fig. 5 (a), with filling factor equals to 0.4, there is a large band gap for *xy* mode between 3$^{rd}$ and 4$^{th}$ band. Through removing rods in honeycomb lattice, a phononic graphene with periodic arrangement of cavities leads to extra bands in the bandgap. For those bands generated from cavity modes, the tight-binding description is applicable since every mode is sufficiently localized in a cavity [23]. To find the counterpart of classical vibration, we only take p-orbital into account because its vibrating direction corresponds to the motion of mechanical oscillators. Also the hopping of the tightly confined state to its neighbour does not introduce the change of polarization from L to T and vice versa. In accordance with above description, the corresponding equation is naturally identical to equation (5). The cavity mode in the band gap is plotted in Fig. 5(b). Its appearance matches the case $C_L \sim C_T$ in mechanical graphene. To further verify its feature of SOC, Fig. 5(c) illustrates the pseudospin texture derived from equation (6) at K point in accordance with the fitting central frequency $\omega_0 = 3322.7$ and two hopping parameters $t_L = 3.96$, $t_T = 3.24$ in the unit of hertz, which gives similar spin distribution to that of spring-mass system. An advantage of cavity-phononic crystals to the spring-mass systems is that they can provide very close L and T elastic constants. Consequently, an effective SOC theory in Sec. II can be utilized and we predict that ESHE at K(K') point is available for such cavity-phononic crystals.

## V. CONCLUSION

In conclusion, the effective spin-orbit coupling in mechanical graphene has been studied. We show that the band structure exhibits a Dresselhaus type of SOC caused by LT splitting. Under the condition of $C_L \sim C_T$, the elastic SHE arising from SOC leads to interesting spin texture that shows up in the field distribution evolution in time. The numerical result verifies the existence of elastic SHE at $\Gamma$ and K(K') point, which also fits the result given in spintronics and photonics. Surprisingly, in practical condition $C_L \gg C_T$, elastic SHE still appears in spite of no perfect field inversion at K(K')point. In the final part we introduce a cavity-phononic crystal that is equivalent to the proposed spring mass system.

**REFERENCE**

[1] J. Sinova, S. Valenzuela, J. Wunderlich, C. Back, and T. Jungwirth, *Reviews of Modern Physics*, **87**, 1213. (2015)

[2] Bin Zhou, *Phys. Rev. B*, **81**, 075318. (2010)

[3] M.-H. Liu and C.-R. Chang, *Phys. Rev. B*, **82**, 155327. (2010)

[4] M. Onada, S. Murakami, and N. Nagaosa, *Phys. Rev. Lett*, **93**, 083901. (2004)

[5] K. Bliokh and Y. Bliokh, *Phys. Rev. Lett*, **96**, 073903. (2006)

[6] X. Yin, Z. Ye, J. Rho, Y. Wang, and X. Zhang, *Science*, **339**, 1405. (2013)

[7] A. Khanikaev, S, Mousavi, W.-K Tse, M. Kargarian, A. MacDonald, and G. Shvets, *Nat. Mater.* **12**, 233, (2013)

[8] C. Kane and E. Mele, *Phys. Rev. Lett*, **95**, 226801. (2005)

[9] K. Bliokh and V. Freiliker, *Phys. Rev. B*, **74**, 174302. (2006)

[10] A. Kavokin, G. Malpuech, and Mikihail Glazov, *Phys. Rev. Lett.* **95**, 136601. (2005)

[11] C. Leyder, M. Romanelli, J. Karr, E. Giacobino, T. Liew, M. Glazov, A. Kavokin, G. Malpuech, and A. Bramati. *Nat. Phys.* **3**, 628. (2007)

[12] V. Sala, D. Solnyshkov, I. Carusotto, T. Jacqmin, A. Lemaître, H. Terças, A. Nalitov, M. Abbarchi, E. Galopin, I. Sagnes, J. Bloch, G. Malpuech, and A. Amo. *Phys. Rev. X,* **5**, 011034. (2015)

[13] A. Nalitov, G. Malpuech, H. Terças, and D. Solnyshkov. *Phys. Rev. Lett.* **114**, 026803. (2015)

[14] S. Murakami, N. Nagaosa, and S. -C. Zhang, *Science*, **301**, 1348, (2003)

[15] J. Sinova, D.Culcer, Q. Niu, N. Stinitsyn, T. Jungwirth, and A. MacDonald. *Phys. Rev. Lett*. **92**, 126603. (2004)

[16] Y. Kato, R. Myers, A. Gossard, and D. Awschalom. *Science*, **306**, 1910. (2004)

[17] L. Zhang and Q. Niu. *Phys. Rev. Lett.* **115**, 115502 (2015)

[18] T. Kariyado and Y. Hatsugai. *Sci. Rep.* **5**, 18107. (2015)

[19] Y. -T. Wang, P. -G. Luan, and S. Zhang. *New J. Phys.* **17**, 073031. (2015)

[20] R. Saito, G. Dresselhaus, and M. Dresselhaus. *Phys. Rev. B*, **61**, 2981. (2000)

[21] S. Konschuh, M. Gmitra, D. Kochan, and J. Fabian. *Phys. Rev. B*, **85**, 115423. (2012)

[22] Z. Qiao, H. Jiang, X. Li, Y. Yao, and Q. Niu. *Phys. Rev. B*, **85**, 115439. (2012)

[23] K. Fang, Z. Yu, and S. Fan. *Phys. Rev. B,* **84**, 075477. (2011)

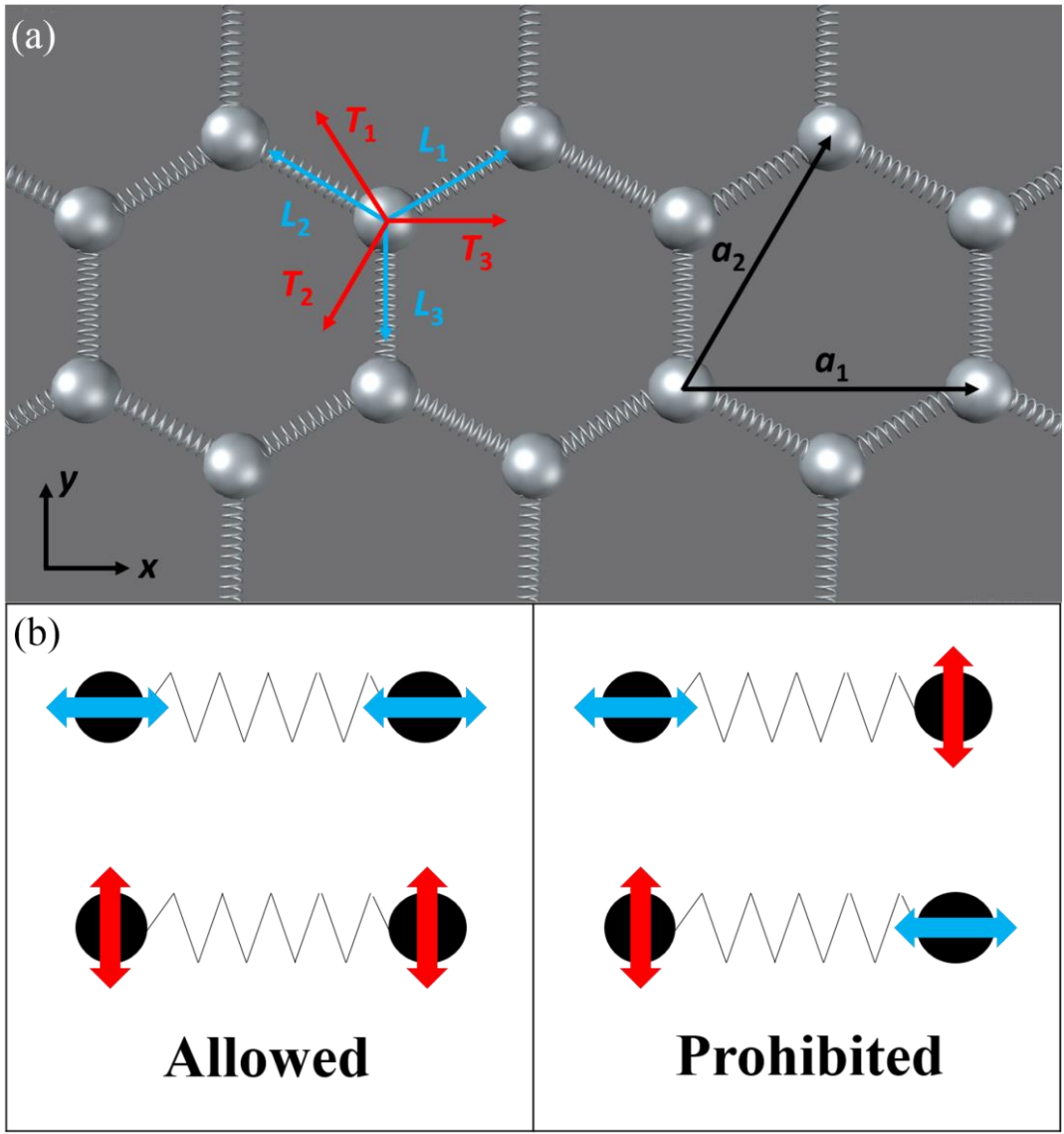

Figure 1: (a) A part of the mechanical graphene made by soft springs and rigid particles. (b) A schematic sketch of the allowed and prohibited vibrating mode transfer.

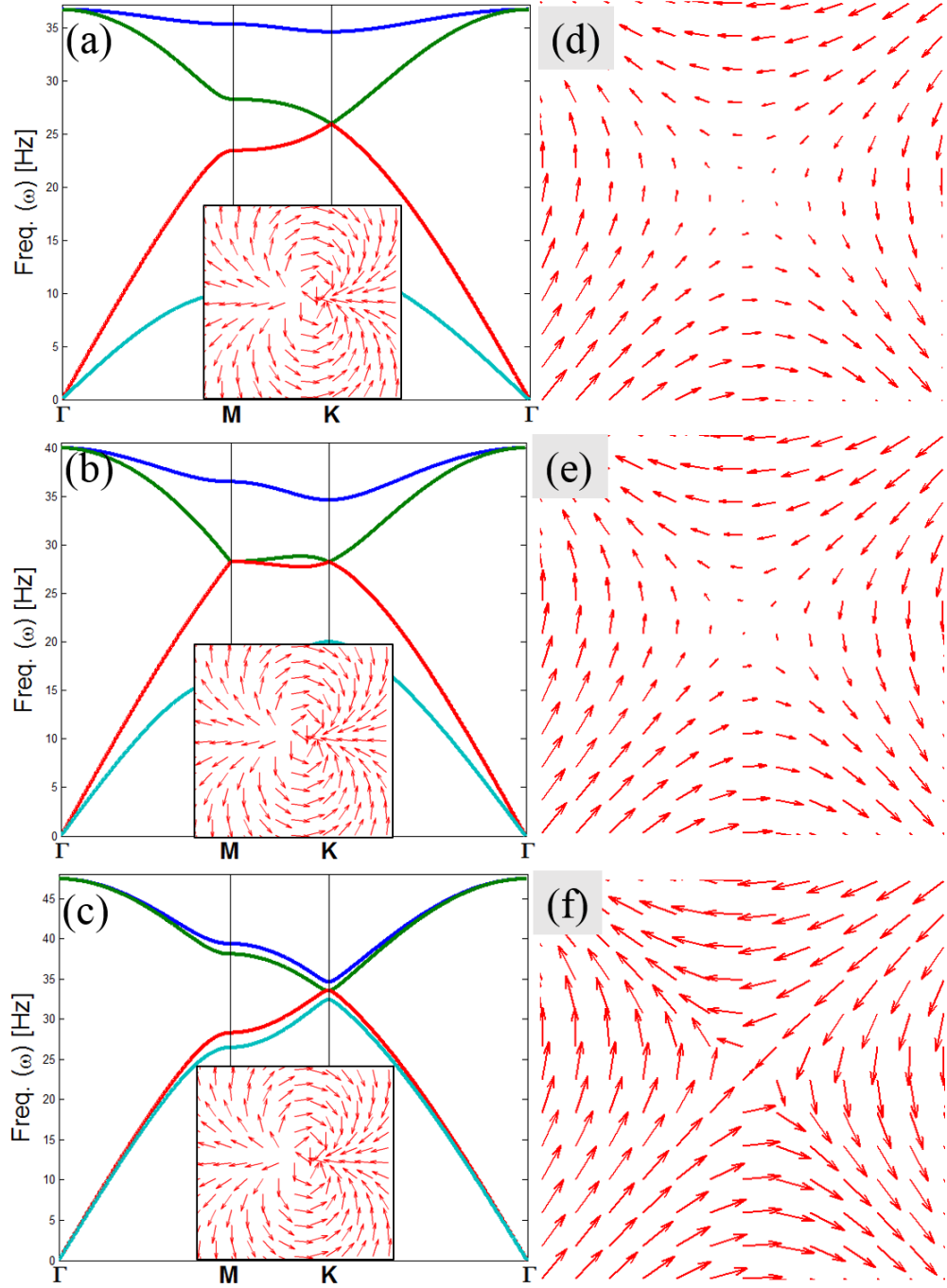

Figure 2(a)-(c): The phonon dispersion for $C_T = C_L/8$, $C_L/3$, and $7C_L/8$ ($M = 10g$ and $C_L = 4$ N/m). The insets are projected spin texture around $\Gamma$ point. (d)-(f) are projected spin textures of third band for corresponding transverse elastic constants in the vicinity of K point. The length of arrows in spin textures indicates the strength of effective magnetic fields.

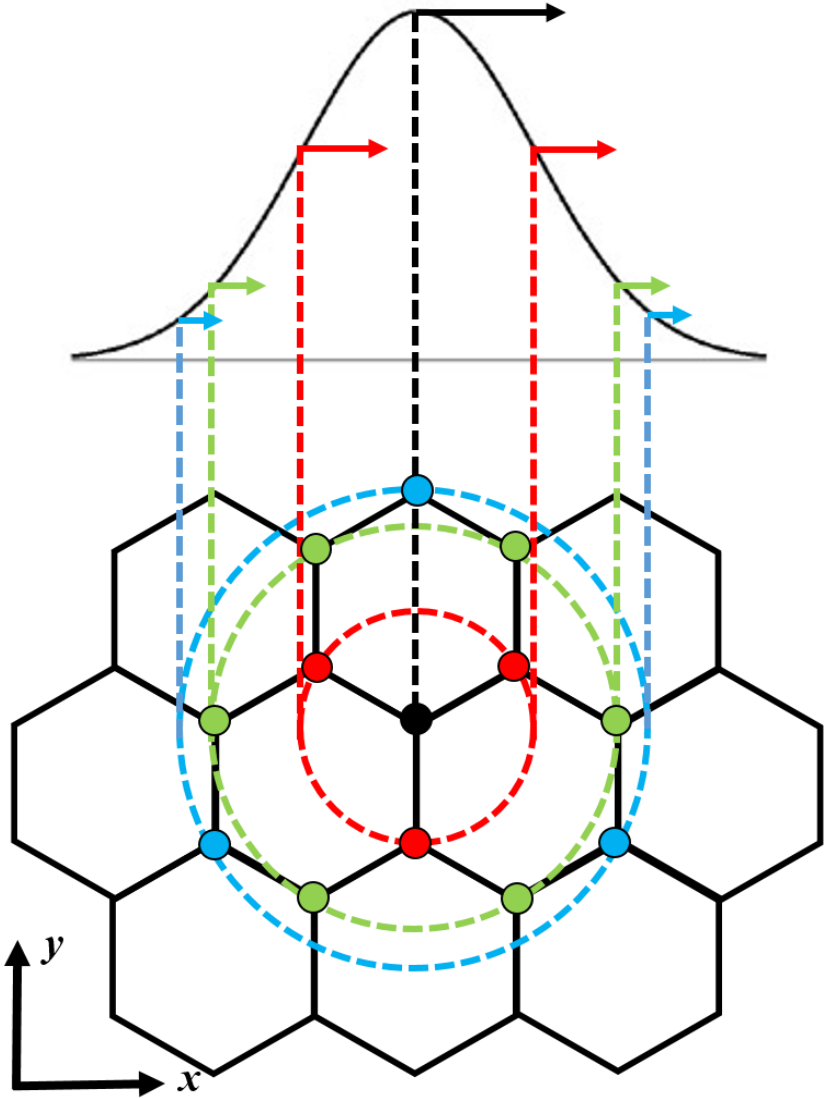

Figure 3: A schematic sketch of the initial displacement distribution. A Gaussian pulse $A_0 \exp\left[-(\mathbf{r}-\mathbf{R}_0)^2/\sigma^2\right]\exp\left[-t^2/\tau^2\right]\exp\left[i(\mathbf{k}\cdot\mathbf{r}-\omega t)\right]$ is applied to the middle of the system, where it has a spatial deviation $\sigma = 6a$ and time deviation $\tau = 25$ ms. All the displacements are along $x$ direction and the length of arrows denotes the corresponding displacements in the Gaussian pulse.

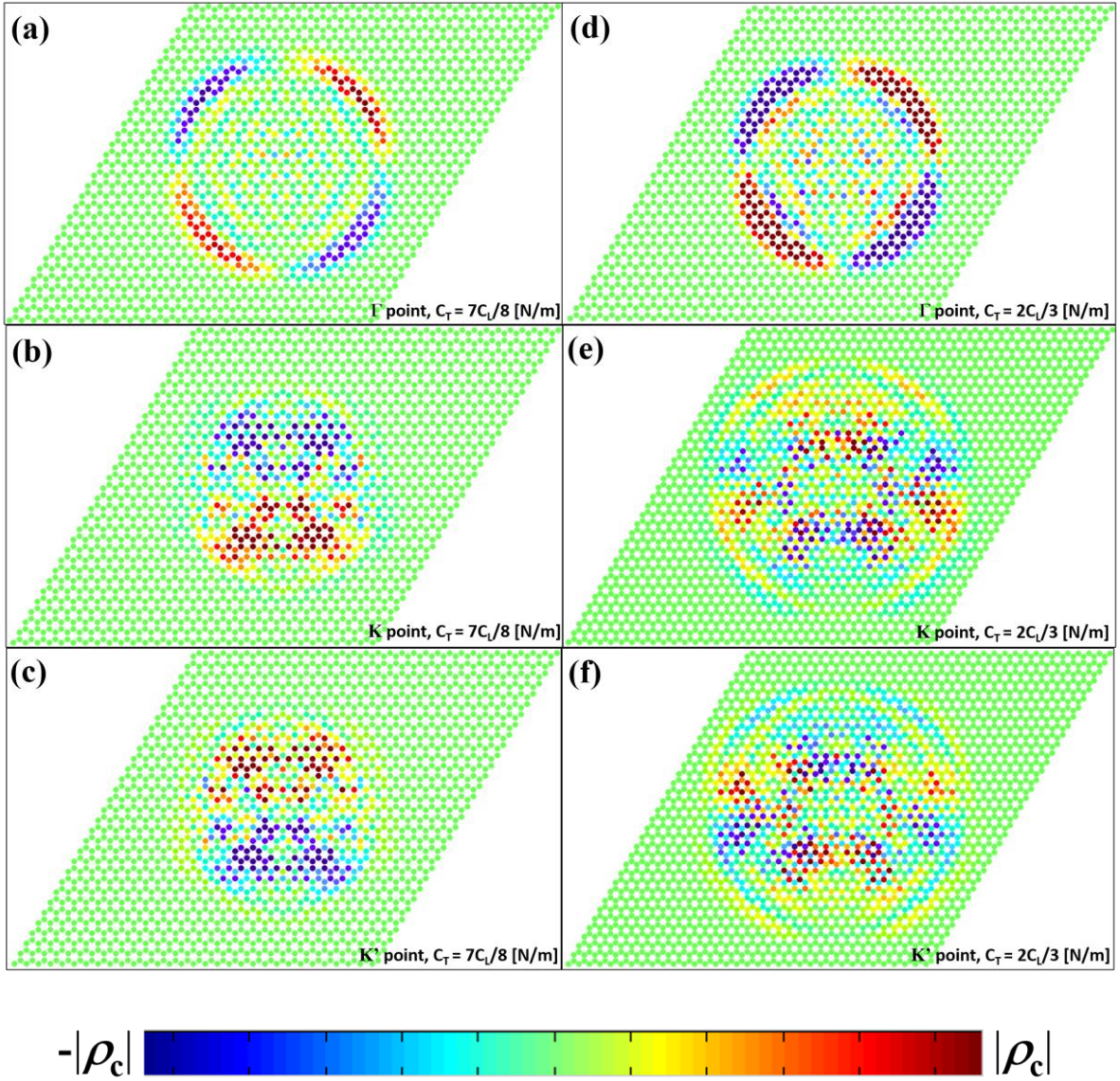

Figure 4(a)-(c): ESHE for $C_T = 7C_L/8$ N/m at $\Gamma$, K and K' point. There are four spin polarizations in (a), two in (b) and (c) propagates separately in distinct orientations. Also (b) and (c) shows a clear inverted field pattern. (d)-(f): ESHE for $C_T = 2C_L/3$ N/m at $\Gamma$, K and K' point. It is clear to observer the existence of splitting of spin-envelope propagation. Despite the compliance of field inversion, (e) and (f) show a mixed spin field pattern since the violation of Dresselhaus SOC. Note that for observing a clear evolution for K and K' points, we increase the number of lattice in the system to 35×35 and add the time step up to 150 in Fig. 4(e) and (f).

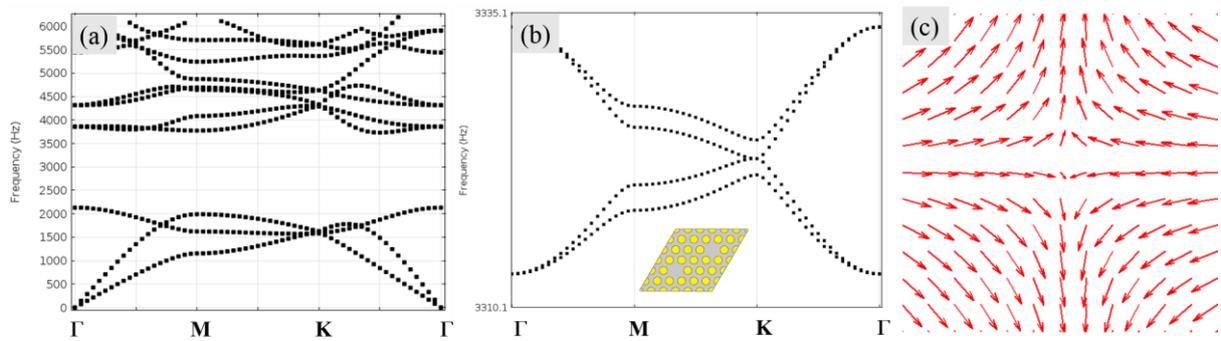

Figure 5: (a) The phononic band structure for tungsten carbide rods embedded in an aluminium background with lattice constant = 1 m. (b) Cavity-phononic crystal shows the dispersion being similar to Fig. 2(a). The inset demonstrates the geometry of a unit cell. (c) Spin texture of Dresselhaus SOC around K point. The orientation discrepancy from Fig. 2(d) is caused by the choice of primitive translation vectors.